# Pure Even Harmonic Generation from Oriented CO in Linearly Polarized Laser Fields

Hongtao Hu,[1,2] Na Li,[1] Peng Liu,[1, 3,]* Ruxin Li,[1,3,†] and Zhizhan Xu[1,‡]

[1]*State Key Laboratory of High Field Laser Physics, Shanghai Institute of Optics and Fine Mechanics, Chinese Academy of Sciences, Shanghai 201800, China*

[2]*University of Chinese Academy of Sciences, Beijing 100049, China*

[3]*Collaborative Innovation Center of IFSA (CICIFSA), Shanghai Jiao Tong University, Shanghai 200240, China*

*\* peng@siom.ac.cn*

*†ruxinli@mail.shcnc.ac.cn*

*‡zzxu@mail.shcnc.ac.cn*

The first high harmonic spectrum, containing only the odd orders, was observed in experiments 30 years ago (1987). However, a spectrum containing pure even harmonics has never been observed. We investigate the generation of pure even harmonics from oriented CO molecules in linearly polarized laser fields employing the time-dependent density-functional theory. We find that the even harmonics, with no odd orders, are generated with the polarization perpendicular to the laser polarization, when the molecular axis of CO is perpendicular to the laser polarization. Generation of pure even harmonics reveals a type of dipole acceleration originating from the permanent dipole moment. This phenomenon exists in all system with permanent dipole moments, including bulk crystal and polyatomic molecules.

High-order harmonic generation (HHG), the most prosperous physical phenomenon of intense field laser-matter interactions, has irreplaceable applications in many research fields: coherent X-ray generation [1–3], chemical reaction tracing [4], molecular orbital imaging [5–7], and attosecond science [8–12], although it has been only three decades since its first experimental observation [13]. The semiclassical three-step model [14,15] in the strong-field approximation [16] provides an easy-to-understand explanation, in which HHG spectra encode the information of the interaction between the ground state of an atom or molecule and the continuum wave function of electrons within intense laser fields.

Normally, HHG only contains odd-order harmonics [17] for the orbital symmetry along the direction of laser polarization. Both even and odd harmonics can be obtained from the oriented asymmetric molecules like CO [18–20] and COS [21], or from symmetric molecules (atoms) in asymmetric laser fields [7,22]. It is a general understanding that HHG of only even orders cannot be generated, since the odd harmonics can always be allowed, even under a symmetry-breaking configuration. In this work, we find that pure even harmonics are produced from oriented CO molecules when the molecular axis is perpendicular to the direction of a linearly polarized intense field. This phenomenon calculated by time-dependent density-functional theory (TDDFT) is confirmed by the plane-wave approximation [5] and the interference of dipole acceleration. The plane-wave approximation explains the polarization properties of the spectra. In addition, the dipole acceleration interference explains why the spectrum contains pure even harmonics. The phenomenon is not only valid for the case of θ = 90°, but also exists in molecules with a finite angle distribution. The analysis of pure even harmonic generation (PEHG) reveals the interaction between a system with a permanent dipole moment and intense laser field. This phenomenon exists in all systems with a permanent dipole moment such as bulk crystal and polyatomic molecules. It will have wide range of applications in an intense laser field.

Quantum theory is a precise tool with which to explain the physics at the atomic and molecular levels [23,24]. Even for the simplest diatomic molecule, solving a three-dimensional (3D) Schrödinger equation incurs an unacceptable computation cost. TDDFT [25–28], the fundamental variation of which is electron density, greatly reduces the computation cost and makes it possible to deal with the multi-electron system. TDDFT is performed by solving the time-dependent Kohn-Sham (TDKS) equation for the orbital wavefunctions $\psi_i(\mathbf{r},t)$ (atomic units are used hereinafter),

$$i\frac{\partial}{\partial t}\psi_i(\mathbf{r},t) = \left\{-\frac{\nabla^2}{2} + V_{\text{eff}}[\rho(\mathbf{r},t)]\right\}\psi_i(\mathbf{r},t)$$

, (1)

where $\rho(\mathbf{r},t)$ is the electron density, given by $\rho(\mathbf{r},t) = \sum_i |\psi_i(\mathbf{r},t)|^2$. The effective potential $V_{\text{eff}}[\rho(\mathbf{r},t)]$ is a function of $\rho(\mathbf{r},t)$,

$$V_{\text{eff}}[\rho(\mathbf{r},t)] = V_{\text{ion}}(\mathbf{r}) + V_{\text{Hartree}}[\rho(\mathbf{r},t)] + V_{\text{xc}}[\rho(\mathbf{r},t)] + V_{\text{laser}}(t) \quad (2)$$

The first term, $V_{\text{ion}}(\mathbf{r})$, is the ion-electron interaction potential. Instead of the real Coulomb potential, the norm-conserving, non-local Troullier-Martines pseudopotential [29] with the Kleinman-Bylander [30] form is used to describe the interaction. 10 electrons (four electrons from C and six electrons from O) are involved in our simulation. The bond length of CO is fixed at 2.14 a.u. (1 a.u. = 0.0529 nm). This nuclear-fixed assumption is reasonable since our laser-pulse duration of 27 fs is short enough that we can ignore the nuclear motion. The second term, $V_{\text{Hartree}}[\rho(\mathbf{r},t)] = \int \frac{\rho(\mathbf{r}',t)}{|\mathbf{r} - \mathbf{r}'|} d\mathbf{r}'$, is the time-dependent Hartree potential, describing the interaction between electrons. The third term, $V_{\text{xc}}[\rho(\mathbf{r},t)]$, is the exchange-correction potential, comprising all the non-trivial many-body effects. We use the generalized-gradient approximation (GGA) of Leeuwen and Baerends [31,32] to describe this potential, and its validity in the simulation of HHG has been verified by many groups [33–36]. The last term, $V_{\text{laser}}(t) = z \cdot E_0 f(t) \sin(\omega t)$, describes the potential induced by the intense laser field at length gauge using the dipole approximation. The polarization of the laser field, propagated in the $y$-axis direction, is set in the direction along the $z$ axis, and the intensity is $1.5 \times 10^{14}$ W/cm$^2$ at the center wavelength of 800 nm. The trapezoidal envelope $f(t)$ ramps linearly during the 1-cycle, stays constant for the 8-cycle, and decays to zero again in the 1-cycle.

The TDKS equation is solved and propagated using the software package OCTOPUS [37–40], in real-space grids with a time step of 0.02 a.u. (1 a.u. = 0.024 fs). We perform the simulation in a large box, the 3D length (space grid) of which is 50 a.u. (0.38 a.u.), 50 a.u. (0.38 a.u.), and 80 a.u. (0.38 a.u.), respectively, to ensure that the returning electron wave functions do not collide with the boundary. To avoid the reflection of ionized electrons from the boundary of the box, a complex absorbing potential [41] is used,

$$V_{absorb}(x) = \begin{cases} 0 & \text{, if } 0 < x < x_{max} \\ i\eta \sin^2\left[\dfrac{(x - x_{max})\pi}{2L}\right] & \text{, if } x_{max} < r < x_{max} + L \end{cases}, \quad (3)$$

where $\eta$ ($= -0.8$ a.u.) is the absorbing potential height, and $L$ ($= 10$ a.u.) is the absorbing length. The HHG spectrum is defined as the absolute square of the Fourier transform of the time-dependent dipole acceleration [42–46],

$$H(\omega) = \left| \int dt \, e^{-i\omega t} \frac{d^2}{dt^2} \left\langle \Psi(t) \middle| \hat{e} \cdot D \middle| \Psi(t) \right\rangle \right|^2. \quad (4)$$

Fig. 1(a) shows the spectrum (dark-yellow line) in the $z$-axis direction for the case $\theta = 0°$ that has been observed experimentally by many groups [18,19]. As Figs. 1(b) and 1(c) show, pure odd (even) harmonics is only generated in the $z$-axis ($x$-axis) direction. Therefore, one can obtain pure even harmonics from the $x$-axis component using the polarization technique [47]. Considering the two facts that the even-to-odd intensity ratio (EOIR) of most orders in this work is greater than 0.04 and an EOIR as low as 0.01 has been measured experimentally [48], the even harmonics discussed here is detectable. Also, a similar phenomenon in HHG from bulk material has been reported [49], proving the validity of our work. Since the even harmonics originates from the permanent dipole moment, we suspect that the efficiency (or EOIR) is related to the value of the permanent dipole moment.

Multiple orbital effects have been found in the HHG of aligned molecules [17]. We first investigate whether this effect also plays a key role in the oriented CO. The molecular populations of HOMO, HOMO-1, and HOMO-2 (where HOMO denotes highest occupied molecular orbital) as a function of time is investigated (not shown here), from which we find that the HOMO has an ionization rate of nearly 5% at the end of laser pulses, while HOMO-1 is ionized less than 0.3% and HOMO-2 is almost un-ionized. According to the three-step model [14–16], a portion of the electrons recombine with the ground state emitting the high-order harmonics and the others are ionized. This means that the greater the ionization of the orbital, the more the orbital contributes to the HHG. Since the HOMO of CO has an ionization rate higher than HOMO-1 (HOMO-2) by 1 order of magnitude, it is evident that HOMO dominates the contribution to HHG. Therefore, the multiple-orbital effect can be neglected here, and we only need to consider the contribution of HOMO in the following discussions.

In the plane-wave approximation [5,50], we can investigate the polarization properties of harmonics. For example, whether the HHG can be produced with the polarization perpendicular to the laser polarization when the CO molecular axis is perpendicular to the polarization of the laser. The calculated $|d_z(\omega, \theta = 90^\circ)|$ (dashed blue line) and $|d_x(\omega, \theta = 90^\circ)|$ (solid red line) are shown in Fig. 2. The figure clearly shows that the harmonics in the $x$-axis direction can be generated when the molecular axis is perpendicular to the laser polarization ($\theta = 90°$), the reason being that the asymmetry of $\varphi_{\text{HOMO}}(x, y, z, \theta = 90^\circ)$ is in the $x$-axis direction. In the case of $\theta = 0°$, the harmonics in the $x$-axis direction disappears for the HOMO symmetry in the $x$-axis direction (as the black dotted line shows in Fig. 2). One can demonstrate that if $\varphi(x, y, z) = \varphi(-x, y, z)$, then $d_x(\omega)$ is zero [51]. Therefore, if $\varphi(x, y, z) \neq \varphi(-x, y, z)$, harmonics will be generated in the $x$-axis direction.

Here, we discuss why the spectrum in the $x$-axis direction contains pure even harmonics. As Fig. 3 shows, the harmonics of the $z$ axis [blue curve, Fig. 3(a)] has an opposite phase at every half-cycle due to the change of sign of the driving laser field [52]. The expression of interference of two spectra separated by a half-cycle in the $z$-axis direction is written as $S_z - S_z e^{iN\omega\Delta t}$ [52]. $S_{z(x)}$ is the HHG spectrum in the $z$-axis ($x$-axis) direction, $\omega$ the frequency of the laser field, $N$ the harmonic order, and $\Delta t = T/2 = \pi/\omega$ the time separation. While the interference expression of the $x$ axis is written as $S_x + S_x e^{iN\omega\Delta t}$, the harmonics in the $x$-axis direction [red curve, Fig. 3(b)], perpendicular to the polarization of the laser field, have only a positive phase every half-cycle. To the best of our knowledge, this type of dipole acceleration originating from the permanent dipole moment has never been investigated. Substituting $\Delta t = T/2 = \pi/\omega$ into the expressions, one can obtain $S_z(1 - e^{iN\pi})$ and $S_x(1 + e^{iN\pi})$, or, more straightforwardly,

$$S_z^{\text{interference}} = \begin{cases} 2S_z, & N = \text{odd} \\ 0, & N = \text{even} \end{cases}$$

(5)

and

$$S_x^{\text{interference}} = \begin{cases} 0, & N = \text{odd} \\ 2S_x, & N = \text{even} \end{cases}.$$

(6)

This undoubtedly means that the spectrum of the $z$ axis ($x$ axis) contains only odd (even) harmonics. The periodicity (every half-cycle) of the dipole acceleration in the $x$-axis direction comes from the symmetry of HOMO in the $z$-axis direction, $\varphi_{\text{HOMO}}(x, y, z) = \varphi_{\text{HOMO}}(x, y, -z)$.

From the discussions above, it is clear that if the laser is polarized linearly in the $z$-axis direction, pure even harmonic generation requires two conditions: (i) $\varphi(x, y, z) \neq \varphi(-x, y, z)$, and (ii) $\varphi(x, y, z) = \varphi(x, y, -z)$. Condition (i) guarantees that there is a net dipole in the *x*-axis direction (see the derivation of formula A in the Supplemental Material). Condition (ii) guarantees that dipole acceleration of the *x* axis has same phase [the red curve shown in Fig. 3(b)] at successive half-cycles of the driving laser field, which is the nature of pure even harmonic generation. Actually, condition (i) requires that the molecule be asymmetric or possess a permanent dipole moment, and condition (ii) requires that the permanent dipole moment is oriented in the direction perpendicular to the polarization of the laser field. The polarization properties of even (odd) harmonics polarized in the *x*-axis (*z*-axis) direction still needs experimental verification.

Although the case $\theta = 90°$ contains all that is principally necessary to generate pure even harmonics and is easy to analyze, the molecules can only be oriented in a finite degree of a typical axis in experiments. Fortunately, this phenomenon also exists in molecules with a finite angle distribution. The harmonic intensity of molecules oriented along the *x* axis is shown in Fig. 4. A distribution of $\cos^2\alpha$ is used in the simulation because the degree of orientation of this angle distribution (0.83) is very close to the experimental value (0.73–0.82) of Ref. [53], which is the highest degree of completely field-free orientation reported so far. The spectrum shown in Fig. 4 is the Fourier transform of the total dipole acceleration. (See the Supplemental Material for the calculation of degree of orientation and total dipole acceleration.)

As shown in Fig. 4, the spectrum of the *x* axis (red solid circles) contains pure even harmonics, just like the case of $\theta = 90°$ [red line in Fig. 1(c)]. The reason for this is the interference between the dipole acceleration of two angles that possess symmetry properties along the *x* axis [for example, $\theta = 70°$ ($\alpha = -20°$) and $\theta = 110°$ ($\alpha = 20°$)]. As shown in Figs. 5(a) and 5(b), the only difference between a dipole acceleration of θ = 70° and one of θ = 110° in the *x*-axis direction is a phase delay of π. Thus, the sum spectra of θ = 70° and 110° [Fig. 5(c)] can be written as

$$S_x^{\theta=70^o} + S_x^{\theta=110^o} = S_x^{\theta=70^o} + e^{-iN\pi} S_x^{\theta=70^o} = S_x^{\theta=70^o}(1+e^{-iN\pi}) , \qquad (7)$$

where *N* is the order of harmonics. This formula, just like formula (6), leads to the spectrum of the *x* axis containing pure even harmonics. For the case of the *z* axis, the sum spectra of θ = 70° and 110° [Fig. 5(f)] can be written as

$$S_z^{\theta=70^o} + S_z^{\theta=110^o} = S_z^{\theta=70^o} - e^{-iN\pi} S_z^{\theta=70^o} = S_z^{\theta=70^o}(1-e^{-iN\pi}) , \qquad (8)$$

since the dipole accelerations of θ = 70° and 110° have opposite directions in addition to the phase delay of π [as shown in Figs. 5(d) and 5(e)]. Therefore, the spectrum in the *z*-axis direction contains only odd orders (blue solid squares in Fig. 4).

In summary, we generate the pure even harmonics from oriented asymmetric molecules (with and without an angle distribution effect) in linearly polarized laser fields theoretically, which is an important step in completing the HHG theory. We also provide a simple and reliable method, in principle, of completely separating odd and even harmonics, which is a potentially useful tool in high harmonic spectroscopy. Finally, we invigorate the investigation of a interaction in a system with a permanent dipole moment and intense laser field. This phenomenon exists in all system with a permanent dipole moment, such as bulk crystal and polyatomic molecules, and will have a wide range of applications in the intense laser field.

ACKNOWLEDGMENT

This work was supported by the National Natural Science Foundation of China (Grant Nos. 11274326, 61521093, 61405222, and 11127901).

FIG. 1. (a)–(c) are HHG spectra of oriented CO. As drawn in (d), θ is the angle between the molecular axis (dashed black line) and the driving laser polarization (red double-arrow). (e) and (f) show the cases θ = 0° and 90°, respectively. The three bold, dark-yellow double arrows shown in (e) and (f) represent the polarization of HHG corresponding to the spectrum of (a)–(c). Spectrum in (c) contains pure even-order harmonics.

FIG. 2. Modulus of dipole in the $z$- and $x$-axis directions for the parallel (θ = 0°) and perpendicular (θ = 90°) cases calculated using plane-wave approximation. The brown double arrow represents the HHG polarization.

FIG. 3. (a) and (b) shown the dipole acceleration of the $z$ axis (blue curve) and $x$ axis (red curve) at the case of θ = 90°, respectively. To show the high oscillations more clearly, the fundamental laser field is subtracted from the $z$-axis dipole acceleration. The plus sign (+) means that the harmonic radiation has positive phase. The minus sign (−) means the harmonic radiation has negative phase, which is opposite the positive phase. The harmonics of the $z$ axis emitted at successive half-cycles of the driving laser field has opposite phase, while the $x$-axis harmonics has only positive phase.

FIG. 4. Harmonic intensity of molecules oriented along the $x$ axis for the polarization along $z$- (odd orders, blue solid squares) and $x$-axis (even orders, red solid circles) directions. The dark-yellow open triangles represent the harmonic intensity in the $x$-axis direction of θ = 90° [as shown in Fig. 1(c)]. As the inset shows, the molecule is oriented along the $x$ axis with an angle distribution of $\sin^2\theta$ (= $\cos^2\alpha$). α (= θ−90°) is the angle between the molecular axis and the $x$ axis.

FIG. 5. (a) Dipole acceleration in $x$-axis direction of θ = 70° (blue line). (b) Dipole

acceleration in *x*-axis direction of θ = 110° (red line). (c) Sum of (a) and (b), θ = 70° + θ = 110° (dark-yellow line). (d)–(f) Same as (a)–(c), but for the *z* axis. The green arrows in (a) and (b) show that the dipole acceleration of θ = 70° in the x-axis direction is the same as that of θ = 110°, except for a time delay of 0.5 cycle or a phase delay of π. The purple arrows in (d) and (e) show that the dipole acceleration of θ = 70° and 110° have opposite directions in addition to the phase delay of π.

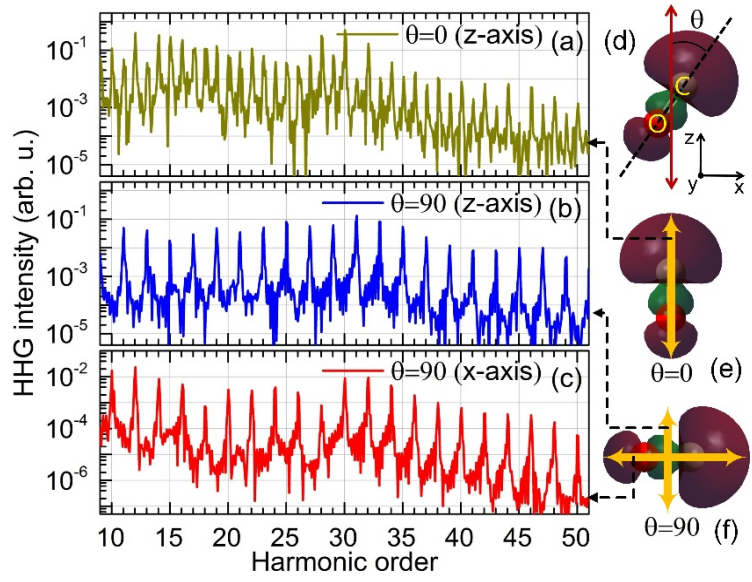

FIG. 1.

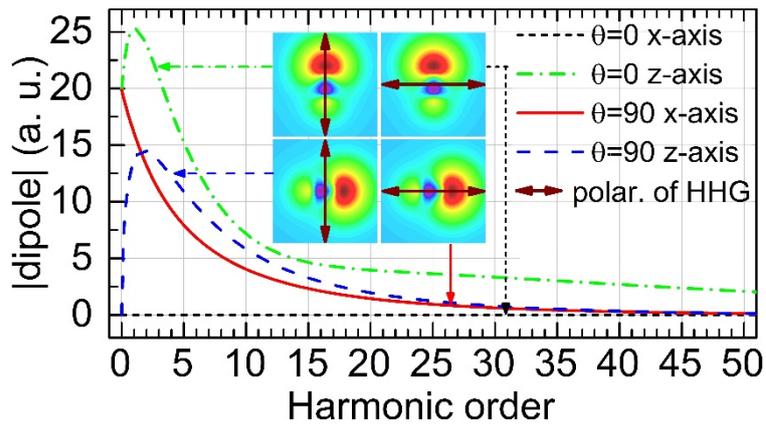

FIG. 2.

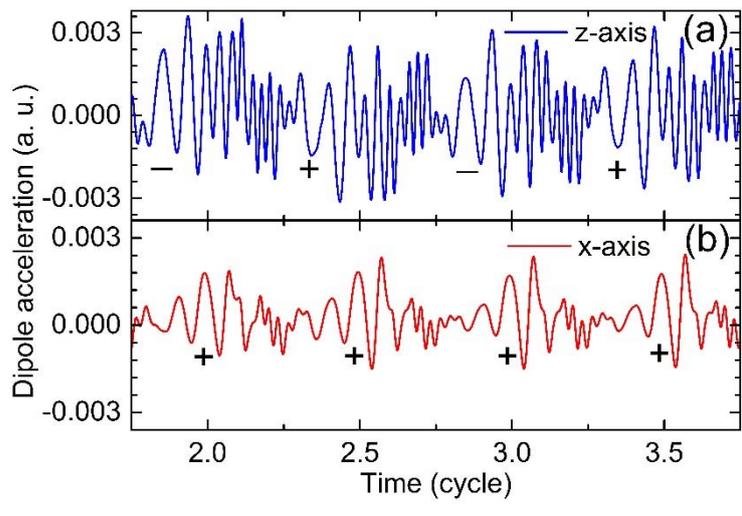

FIG. 3.

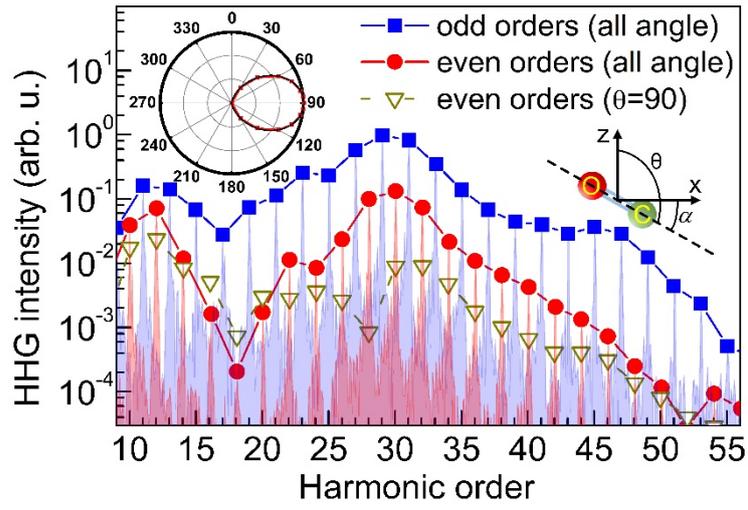

FIG. 4.

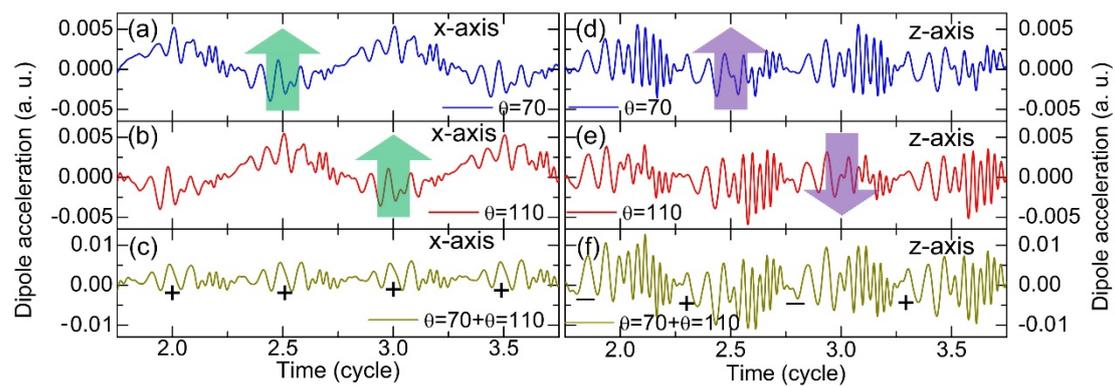

FIG. 5.